\documentclass[12pt]{article}
\usepackage[english]{babel}
\usepackage{graphicx}
\usepackage{amsfonts}
\usepackage{amsmath}
\usepackage{amssymb}
\usepackage{amsthm}
\usepackage{bbold}

\usepackage{natbib}
\usepackage{float}

\usepackage{subfig}
\usepackage{longtable}
\usepackage{pgfplots}
\pgfplotsset{compat = newest}

\usepackage{enumitem}

\allowdisplaybreaks[2]

\DeclareMathOperator*{\argmax}{arg\,max}

\theoremstyle{plain}
\providecommand{\propertyname}{Property}

\theoremstyle{plain}\newtheorem{assumption}{\protect\assumptionname}
\providecommand{\assumptionname}{Assumption}

\theoremstyle{plain}
\providecommand{\GEVname}{GEV}

\theoremstyle{plain}
\providecommand{\ARUMname}{ARUM}

\theoremstyle{plain}\newtheorem{thm}{\protect\theoremname}
\providecommand{\theoremname}{Theorem}

\usepackage{color}
\definecolor{DarkBlue}{rgb}{0,0.18,0.55}
\usepackage{hyperref}
\hypersetup{colorlinks=true,citecolor=DarkBlue,filecolor=DarkBlue,linkcolor=DarkBlue,urlcolor=DarkBlue}
\usepackage[font=small,labelfont=bf]{caption}

\begin{document}

\title{Note on solving one-to-one matching models with linear transferable utility\thanks{This research has benefited from discussions with Nikolaj Nielsen. The author is grateful for comments and suggestions by prof. Fedor Iskhakov. The research is funded by the Australian Research Council (ARC) Future Fellowship (grant agreement No. FT180100632: Solving and estimating dynamic models of strategic interaction).}}
\author{Esben Scriver Andersen\thanks{Research School of Economics, HW Arndt Building 25A, College of Business and Economics, The Australian National University, ACT 2601 Australia, \protect\href{mailto:esbenscriver@gmail.com}{esbenscriver@gmail.com}.}}

\maketitle

\begin{abstract}
We derive a system of fixed-point equations for the equilibrium transfers in a class of one-to-one matching models with linear transferable utility. We then show that, when the degree of substitution between alternatives is bounded from above, the derived system of equations constitutes a contraction mapping. As a result, fixed-point iterations are guaranteed to converge to the unique distribution of equilibrium transfers.  
\end{abstract}


\section{Introduction}
In this note, we derive a system of fixed-point equations for equilibrium transfers within a class of one-to-one matching models with linear transferable utility. We demonstrate that this system of equations constitutes a contraction mapping, provided that the own-elasticities of the optimal choice probabilities are bounded from above. Consequently, for this class of matching models, equilibrium transfers can be determined via fixed-point iterations.\footnote{We provide a \href{https://github.com/esbenscriver/SolveOneToOneMatching}{Python} implementation of our algorithm written in JAX.}


Our results are related to \cite{dupuy2014JPE} and \cite{dupuy2022note} who proposed a system of fixed-point equations for equilibrium transfers for a matching model similar to \cite{choo2006marriage}. Hence, restrict the stochastic payoffs to follow the extreme value type-1 (EV1) distribution. \cite{galichon2017iia} discuss the implied IIA property and its limitations of matching models where the stochastic payoffs are EV1 distributed. 

We contribute by proving that in the EV1 case, fixed-point iterations are guaranteed to converge, and we extend this result to any distribution of the stochastic payoffs where the own-elasticites of the resulting optimal choice probabilities for all agents on both sides of the market are bounded from above.

Our proposed system of fixed point equations can be cast as a nested fixed point in a maximum likelihood estimation procedure as proposed by \cite{dupuy2022note} to allow for more general substitution patterns than the logit model. There exists a literature concerning the numerical stability of estimation procedures relying on nested fixed-points. For the interested reader we refer to \cite{Dube2012, Knittel2014, Lee2016}.

By establishing that our system of fixed-point equations is a contraction mapping, we confirm the existence and uniqueness of equilibrium transfers in this class of matching models. These results complement those of \cite{galichon2023cupidsinvisiblehandsocial}, who used convex analysis to prove existence and uniqueness under weaker distributional assumptions for stochastic payoffs.

Matching with transferable utility has been applied in various fields, including marriage market (e.g. \cite{choo2006marriage, ChiapporiEtAl2017}), goods markets (\cite{chiappori2010hedonic}), international trade (\cite{costinot2015}), industrial organization (\cite{bajari2013estimating, fox2018, fox2018unobserved}), and labor market (e.g. \cite{tervio2008, gabaix2008, DupuyTaxation2020, andersenlee2022}). For concreteness, we will focus on matching in the labor market, where we will refer to agents as “workers” and “firms”, and the transfers between agents as "wages".

The remaining of this note is organized as follows. Section \ref{sec_model} presents the class of one-to-one matching models with linear transferable utility that we study. Section \ref{sec_results} states the main theorem of the note and the distributional assumption for the stochastic payoffs. Section \ref{sec_discussion} discusses the implications of the main theorem, and illustrate how to set up a system of fixed-point equations, that is ensured to define a contraction, for the logit, nested logit, and generalized nested logit model. Section \ref{sec_conclude} concludes.


\section{Model}\label{sec_model}
We consider a matching market where firms and workers face a discrete-choice regarding where to work and whom to hire. The model is closed by a set of market-clearing conditions that determine the distribution of equilibrium wages and matches. 

Let $\mathcal{X}$ denote the set of unique types of workers, and let $\mathcal{Y}$ denote the set of unique types of firms. We will assume that the number of types, $\left|\mathcal{X}\right|$
and $\left|\mathcal{Y}\right|$, are finite.

$\mathcal{Y}_0$ denotes the full choice set of the workers. A worker $a$ of type $x$ faces the discrete-choice of not working, $y=0$, or working for one of the $|\mathcal{Y}|$ types of firms, $y=1,...,|\mathcal{Y}|$,
\begin{equation}\label{eq_ARUM_worker}
    \max_{y \in \mathcal{Y}_0} \bigg\{ v_{xy}^X + \sigma_x^X \varepsilon_{axy}^X \bigg\}.
\end{equation}
$v_{xy}^X$ and $\varepsilon_{axy}^X$ are the deterministic and stochastic parts of the payoff function, and $\sigma_x^X > 0$ is the scale parameter. If the worker chooses not to work, they derive a deterministic payoff of zero. When a worker of type $x$ chooses to work for a firm of type $y$, their deterministic payoff is given by the sum of the utility term, $\beta_{xy}^X$, and the wage, $w_{xy}$
\begin{align}
    v_{x0}^X &= 0, \\
    v_{xy}^X &= \beta_{xy}^X + w_{xy} \; \forall \; y \in \mathcal{Y}. \label{eq_payoff_workers}
\end{align}
As we assume there exists a continuum of each type of worker, the share of workers of type $x$ matched with a firm of type $y$ is given by the choice probabilities, $p_{xy}^X: \mathbb{R}^{|\mathcal{Y}|} \rightarrow \mathbb{R}$. These are functions of the wages that the workers face, $w_{x \cdot}=(w_{x1},\cdots,w_{x|\mathcal{Y}|})$  
\begin{align}
    p_{xy}^X(w_{x \cdot}) 
    &\equiv Pr\bigg[ y = \argmax_{j \in \mathcal{Y}_0} \bigg\{v_{xj}^X + \sigma_x^X \varepsilon_{axj}^X \bigg\} \Big|w_{x \cdot} \bigg] \; \forall \; y \in \mathcal{Y}_0. \label{eq_ARUM_CCP_worker}
\end{align}
$\mathcal{X}_0$ denotes the full choice set of the firms. Firms are assumed to consist of one potential job. Hence, a firm $b$ of type $y$ faces the discrete-choice of not hiring any worker, $x=0$, or hiring one of the $|\mathcal{X}|$ types of workers, $x=1,...,|\mathcal{X}|$, 
\begin{equation}\label{eq_ARUM_firm}
    \max_{x \in \mathcal{X}_0} \bigg\{ v_{xy}^Y + \sigma_y^Y \varepsilon_{bxy}^Y \bigg\}.
\end{equation}
When a firm of type $y$ chooses to hire a worker of type $x$, its deterministic payoff is given by the productivity term, $\beta_{xy}^Y$, minus the wage, $w_{xy}$
\begin{align}
    v_{0y}^Y &= 0, \\
    v_{xy}^Y &= \beta_{xy}^Y - w_{xy} \; \forall \; x \in \mathcal{X}. \label{eq_payoff_firms}
\end{align}
The share of firms of type $y$ matched with workers of type $x$, $p_{xy}^Y: \mathbb{R}^{|\mathcal{X}|} \rightarrow \mathbb{R}$, is a function of the vector of wages that the firm faces, $w_{\cdot y}=(w_{1y},\cdots,w_{|\mathcal{X}|y})$,  
\begin{align}
    p_{xy}^Y(w_{\cdot y}) 
    &\equiv Pr\bigg[ x = \argmax_{i \in \mathcal{X}_0} \bigg\{v_{iy}^Y + \sigma_y^Y \varepsilon_{biy}^Y \bigg\} \Big|w_{\cdot y} \bigg] \; \forall \; x \in \mathcal{X}_0, \label{eq_ARUM_CCP_firm}
\end{align}

The discrete-choices of workers and firms are connected through the wages that are determined in a competitive equilibrium. Let the mass of workers of type $x$ be denoted by $n_x^X$, and the mass of firms of type $y$ be denoted by $n_y^Y$. The vector of equilibrium wages, $W^*=(w_{11}^*,\cdots,w_{1|\mathcal{Y}|}^*,\cdots,w_{|\mathcal{X}|1}^*,\cdots,w_{|\mathcal{X}||\mathcal{Y}|}^*)$, and the vector of equilibrium matches, $\mu=(\mu_{11},\cdots,\mu_{1|\mathcal{Y}|},\cdots,\mu_{|\mathcal{X}|1},\cdots,\mu_{|\mathcal{X}|\mathcal{Y}|})$, are jointly determined from a set of market-clearing conditions, such that workers' supply of labor equate the firms' demand for labor across all combinations of $x$ and $y$,
\begin{align}\label{eq_market_clearing}
    \mu_{xy}(W^*) = n_x^X p_{xy}^X(w_{x \cdot}^*) = n_y^Y p_{xy}^Y(w_{\cdot y}^*) \; \forall \; (x,y) \in \mathcal{X} \times \mathcal{Y}.
\end{align}
For later use, define the scaled deterministic payoffs as $\Tilde{v}_{xy}^X \equiv v_{xy}^X/\sigma_x^X$ and $\Tilde{v}_{xy}^Y \equiv v_{xy}^Y/\sigma_y^Y$.

\section{Results}\label{sec_results}
We derive, in Appendix \ref{sec_derivation}, from the market-clearing conditions, a system of fixed-point equations for the equilibrium wages\footnote{\cite{dupuy2022note} proposed a system of fixed-point equations of the same form as Eq. \eqref{eq_FixedPointComplicated} in Appendix \ref{sec_derivation}, but for the limited case where the stochastic payoffs are EV1 distributed.}$^,$\footnote{Note that we have implicitly normalized marginal utility of wages to unity. Without this normalization, then Eq. \eqref{eq_payoff_workers} and \eqref{eq_payoff_firms} could be expressed as $v_{xy}^X = \beta_{xy}^X + \eta_x^X w_{xy}$ and $v_{xy}^Y = \beta_{xy}^Y + \eta_y^Y w_{xy}$. Under this specification, the term $\sigma_x^X \sigma_y^Y / (\sigma_x^X + \sigma_y^Y)$ in Eq. \eqref{eq_FixedPoint} changes to $\eta_x^X\sigma_x^X \eta_y^Y\sigma_y^Y / (\eta_y^Y\sigma_x^X + \eta_x^X\sigma_y^Y)$.}
\begin{align}\label{eq_FixedPoint}
    w_{xy}^* = w_{xy}^* + \tfrac{\sigma_x^X \sigma_y^Y}{\sigma_x^X + \sigma_y^Y}\log \left[ \frac{n_y^Y p_{xy}^Y(w_{\cdot y}^*)}{n_x^X p_{xy}^X(w_{x \cdot}^*)} \right] \; \forall \; (x,y) \in \mathcal{X} \times \mathcal{Y}.
\end{align}
Eq. \eqref{eq_FixedPoint} is derived under the assumption that the stochastic payoffs are generalized extreme value distributed. However, since this system of equations simply states that demand must equal supply, it must hold for any distribution. Further, if we change the term $\sigma_x^X \sigma_y^Y / (\sigma_x^X + \sigma_y^Y)$, the solution to the system of equations will remain unchanged
\begin{align}\label{eq_RobustFixedPoint}
    w_{xy}^* = w_{xy}^* + \tfrac{c_{xy}^X\sigma_x^X c_{xy}^Y\sigma_y^Y}{c_{xy}^X\sigma_x^X + c_{xy}^Y\sigma_x^Y} \log \left[ \frac{n_y^Y p_{xy}^Y(w_{\cdot y}^*)}{n_x^X p_{xy}^X(w_{x \cdot}^*)} \right] \; \forall \; (x,y) \in \mathcal{X} \times \mathcal{Y}.
\end{align}

To prove that Eq. \eqref{eq_RobustFixedPoint} defines a contraction mapping we make the following two assumptions.

\begin{assumption}[Full support]\label{assumption_Support}
The stochastic parts of the workers' and firms' payoffs, $(\varepsilon_{x \cdot}^X,\varepsilon_{\cdot y}^Y)$, follow a distribution with full support and are absolutely continuous with respect to the Lebesgue measure.
\end{assumption}

\begin{assumption}[Own-elasticities are bounded]\label{assumption_Elasticity}
     The own-elasticities of the workers' and firms' choice probabilities are bounded for any $W \in \mathbb{R}^{|\mathcal{X}||\mathcal{Y}|}$ such that
    \begin{align*}
        c_{xy}^X \frac{\nabla_{\Tilde{v}_{xy}^X} p_{xy}^X(w_{x \cdot})}{p_{xy}^X(w_{x \cdot})}
        &< 1\; \forall \; (x,y) \in \mathcal{X} \times \mathcal{Y}_0, \\ 
        c_{xy}^Y \frac{\nabla_{\Tilde{v}_{xy}^Y} p_{xy}^Y(w_{\cdot y})}{p_{xy}^Y(w_{\cdot y})}
        &< 1 \; \forall \; (y,x)  \in \mathcal{Y} \times \mathcal{X}_0,
    \end{align*}
    where the constants are strictly positive, $c_{xy}^X > 0$ and $c_{xy}^Y > 0$.
\end{assumption}

Assumption \ref{assumption_Support} is a standard assumption in the discrete-choice literature. It implies that any alternative in the workers' and firms' choice set is chosen with a strictly positive probability. Assumption \ref{assumption_Elasticity} ensures that the own-elasticities of the choice probabilities of the workers and firms are bounded from above. Under Assumption \ref{assumption_Support} and \ref{assumption_Elasticity}, we show, in Appendix \ref{sec_proof}, that Eq. \eqref{eq_RobustFixedPoint} defines a contraction mapping. This is stated below in Theorem \ref{thm_robust}.
\begin{thm}\label{thm_robust}
For the matching model given by Eq. \eqref{eq_ARUM_worker}-\eqref{eq_market_clearing} under Assumption \ref{assumption_Support} and \ref{assumption_Elasticity}, the function $F:\mathbb{R}^{|\mathcal{X}||\mathcal{Y}|} \rightarrow \mathbb{R}^{|\mathcal{X}||\mathcal{Y}|}$ defines a contraction mapping for the wages, $W$, where
\begin{align}
    F(W) = 
    \begin{bmatrix}
        f_{11}(W)& \cdots& f_{1|\mathcal{Y}|}(W)& \cdots& f_{|\mathcal{X}|1}(W)& \cdots& f_{|\mathcal{X}||\mathcal{Y}|}(W)
    \end{bmatrix}^T, \label{eq_FixedPointRobustVector}
\end{align}
and $f_{xy}:\mathbb{R}^{|\mathcal{X}||\mathcal{Y}|} \rightarrow \mathbb{R}$
\begin{align}
    f_{xy}(W)
    =& w_{xy} + \tfrac{c_{xy}^X\sigma_x^X c_{xy}^Y\sigma_y^Y}{c_{xy}^X\sigma_x^X + c_{xy}^Y\sigma_x^Y} \log \left[ \frac{n_y^Y p_{xy}^Y(w_{\cdot y})}{n_x^X p_{xy}^X(w_{x \cdot})} \right] \; \forall \; (x,y) \in \mathcal{X} \times \mathcal{Y}. \label{eq_FixedPointRobustMatch}
\end{align}

\end{thm}

\section{Discussion}\label{sec_discussion}
Theorem \ref{thm_robust} ensures that there exists a unique equilibrium that satisfy
\begin{align*}
    W^* = F(W^*),
\end{align*}
such that demand and supply are equal across all types of matches. This implication of Theorem \ref{thm_robust} is well known, as \cite{galichon2023cupidsinvisiblehandsocial} showed existence and uniqueness of the equilibrium, under weaker distributional assumptions for the stochastic payoffs.

Further, Theorem \ref{thm_robust} implies that we can use fixed-point iterations to solve for equilibrium wages. Hence, starting from any initial guess, $W^0$, and then iteratively update this guess by evaluating $F$
\begin{align}
    W^{k+1} = F(W^k), \label{eq_algorithm}
\end{align}
 is ensured to converge to the unique distribution of equilibrium wages. Note that for a given $W^k$, if firms' demand for the match of type $(x,y)$ exceeds works' supply to the match of type $(x,y)$, then the ratio of demand to supply in Eq. \eqref{eq_FixedPointRobustMatch} is greater than 1 and $\log[\cdot]>0$. In this case, the wage of the match, $w_{xy}$, is adjusted upward, to reduce demand and increase supply. When the demand is below the supply, then the ratio is below 1 and $\log[\cdot]<0$, such that the wage is adjusted downward. The adjustment is repeated iteratively until firms' demand equals workers' supply across all matches in the market.

The term $c_{xy}^X \sigma_x^X c_{xy}^Y \sigma_y^Y / (c_{xy}^X \sigma_x^X + c_{xy}^Y \sigma_y^Y) > 0$ can be seen as an adjustment of the step length (given by the log-ratio of demand and supply). Theorem \ref{thm_robust} implies that, if we can find a set of positive scalars, $(c_{xy}^X,c_{xy}^Y)$, such that the all the own-elasticities of the choice probabilities of both the workers and firms are less than $1 / c_{xy}^X$ and $1 / c_{xy}^Y$, then Eq. \eqref{eq_FixedPointRobustVector}-\eqref{eq_FixedPointRobustMatch} define a contraction mapping.

Our proposed system of fixed-point equations is related to the work of \cite{BLP1995}, who set up a system of fixed-point equations for consumers' unobserved utilities such that observed choice probabilities equate the predicted choice probabilities. Instead, we use the wage to equate the demand and supply. It is well known, that iterating on the system of fixed point equations proposed by \cite{BLP1995} converge linearly, where the rate of convergence is measured by the Lipschitz constant, see e.g. \cite{Knittel2014}. The same holds for our system of fixed point equations, where the Lipschitz constant is the norm of a matrix involving the own- and cross-elasticities with respect to the wage for workers' supply and firms' demand, see Appendix \ref{sec_proof}. When the Lipschitz constant is close to one then the rate of convergence is low. This is for instance the case when the scalars $c_{xy}^X$ and $c_{xy}^Y$ are close to zero.

\subsection{The logit case}
We now show that when the stochastic payoffs are EV1 distributed, Assumption \ref{assumption_Elasticity} must hold. When the stochastic payoffs are EV1 distributed, it is well known that the resulting choice probabilities are given by the logit choice probabilities
\begin{align*}
    p_{xy}^X(w_{x \cdot}) = \frac{\exp{\left(\Tilde{v}_{xy}^X\right)}}{1 + \sum_{j=1}^{|\mathcal{Y}|} \exp{\left(\Tilde{v}_{xj}^X\right)}} \; \forall \; y \in \mathcal{Y}_0,  \\
    p_{xy}^Y(w_{\cdot y}) = \frac{\exp{\left(\Tilde{v}_{xy}^Y\right)}}{1 + \sum_{i=1}^{|\mathcal{X}|}  \exp{\left(\Tilde{v}_{iy}^Y\right)}}  \; \forall \; x \in \mathcal{X}_0, 
\end{align*}
and it follows that the own-elasticities of the choice probabilities are strictly less than one, when $\sigma_x^X > 0$ and $\sigma_y^Y > 0$
\begin{align*}
    \frac{\nabla_{\Tilde{v}_{xy}^X} p_{xy}^X(w_{x \cdot})}{p_{xy}^X(w_{x \cdot})} = 1 - p_{xy}^X(w_{x \cdot}) < 1,  \; \forall \; y \in \mathcal{Y}_0\\
    \frac{\nabla_{\Tilde{v}_{xy}^Y} p_{xy}^Y(w_{\cdot y})}{p_{xy}^Y(w_{\cdot y})} = 1 - p_{xy}^Y(w_{\cdot y}) < 1  \; \forall \; x \in \mathcal{X}_0.
\end{align*}
Consequently, Assumption \ref{assumption_Elasticity} is satisfied, and we can set the scalars in Eq. \eqref{eq_FixedPointRobustMatch} to unity
\begin{align*}
    c_{xy}^X
    &=1 \; \forall \; (x,y) \in \mathcal{X} \times \mathcal{Y}, \\
    c_{xy}^Y
    &=1 \; \forall \; (x,y) \in \mathcal{X} \times \mathcal{Y}.
\end{align*}
It follows from Theorem \ref{thm_robust}, that iterating on Eq. \eqref{eq_algorithm}  is then ensured to converge to the unique distribution of equilibrium wages. 


\subsection{The nested logit case}

Let the $|\mathcal{Y}|$ types of firms belong to $|K|$ mutual exclusive nests, and let $\mathcal{B}_k^Y$ denote the set of firms belonging to nest $k$. Similarly, the $|\mathcal{X}|$ types of workers belong to $|L|$ mutual exclusive nests, and let $\mathcal{B}_{\ell}^X$ denote the set of workers belonging to nest $\ell$. If, without loss of generality, we assume that the outside options are located in their own nests, then the nested logit choice probabilities for workers and firms are given as follows
\begin{align*}
    p_{xy}^X(w_{x \cdot}) 
    &= \frac{\exp{\left(\Tilde{v}_{xy}^X / \lambda_{xm}^X \right)} \left[\sum_{j \in \mathcal{B}_m^Y} \exp{\left( \Tilde{v}_{xj}^X / \lambda_{xm}^X \right)} \right]^{\lambda_{xm}^X - 1}}{1 + \sum_{k=1}^{|K|} \left[\sum_{j \in \mathcal{B}_k^Y} \exp{\left( \Tilde{v}_{xj}^X / \lambda_{xk}^X \right)} \right]^{\lambda_{xk}^X}} \text{ for } y \in \mathcal{B}_m^Y, \\
    p_{xy}^Y(w_{\cdot y}) 
    &=  \frac{\exp{\left(\Tilde{v}_{xy}^Y / \lambda_{ny}^Y \right)} \left[\sum_{i \in \mathcal{B}_n^Y} \exp{\left( \Tilde{v}_{iy}^Y / \lambda_{ny}^X \right)} \right]^{\lambda_{ny}^Y - 1}}{1 + \sum_{\ell=1}^{|L|} \left[\sum_{i \in \mathcal{B}_{\ell}^X} \exp{\left( \Tilde{v}_{iy}^Y / \lambda_{\ell y}^Y \right)} \right]^{\lambda_{\ell y}^Y}} \text{ for } x \in \mathcal{B}_n^X,
\end{align*}
where $\lambda_{xm}^X \in (0,1]$ and $\lambda_{ny}^Y \in (0,1]$ are nesting parameters. The own-elasticities of the nested logit choice probabilities can be expressed as follow
\begin{align*}
    \frac{\nabla_{\Tilde{v}_{xy}^X} p_{xy}^X(w_{x \cdot})}{p_{xy}^X(w_{x \cdot})} 
    &= \tfrac{1}{\lambda_{xm}^X} -  \tfrac{1-\lambda_{xm}^X}{\lambda_{xm}^X} p_{xy|m}^X(w_{x \cdot})  - p_{xy}^X(w_{x \cdot}) \gtrless 1 \text{ for } y \in \mathcal{B}_m^Y, \\
    \frac{\nabla_{\Tilde{v}_{xy}^Y} p_{xy}^Y(w_{\cdot y})}{p_{xy}^Y(w_{\cdot y})} 
    &= \tfrac{1}{\lambda_{ny}^Y} -  \tfrac{1-\lambda_{ny}^Y}{\lambda_{ny}^Y} p_{xy|n}^Y(w_{\cdot y})  - p_{xy}^Y(w_{\cdot y}) \gtrless 1 \text{ for } x \in \mathcal{B}_n^X,
\end{align*}
where $p_{xy|m}^X$ ($p_{xy|n}^Y$) is the conditional probability of choosing alternative $y$ ($x$) given nest $m$ ($n$) was chosen.\footnote{Mathematically $p_{xy|m}^X$ and 
 $p_{xy|n}^Y$ are defined as
\begin{align}
    p_{xy|m}^X(w_{x \cdot}) 
    &\equiv \frac{\exp{\left(\Tilde{v}_{xy}^X / \lambda_{xm}^X \right)} }{\sum_{j \in \mathcal{B}_m^Y} \exp{\left( \Tilde{v}_{xj}^X / \lambda_{xm}^X \right)}} \text{ for } y \in \mathcal{B}_m^Y,   \\
    p_{xy|n}^Y(w_{\cdot y}) 
    &\equiv  \frac{\exp{\left(\Tilde{v}_{xy}^Y / \lambda_{ny}^Y \right)}}{\sum_{i \in \mathcal{B}_n^Y} \exp{\left( \Tilde{v}_{iy}^Y / \lambda_{ny}^X \right)}} \text{ for } x \in \mathcal{B}_n^X,   
\end{align}
} Hence, the own-elasticities of the nested logit choice probabilities are in general not restricted to be less than one. However, \cite{grigolon2014nestedRC} showed that the product of the nesting parameter and the own-elasticity is guaranteed to be less than one
\begin{align*}
    \lambda_{xm}^X \frac{\nabla_{\Tilde{v}_{xy}^X} p_{xy}^X(w_{x \cdot})}{p_{xy}^X(w_{x \cdot})} 
    &= 1 -  (1-\lambda_{xm}^X) p_{xy|m}^X(w_{x \cdot})  - \lambda_{xm}^X p_{xy}^X(w_{x \cdot}) < 1 \text{ for } y \in \mathcal{B}_m^Y,\\
    \lambda_{ny}^Y \frac{\nabla_{\Tilde{v}_{xy}^Y} p_{xy}^Y(w_{\cdot y})}{p_{xy}^Y(w_{\cdot y})} 
    &= 1 -  (1-\lambda_{ny}^Y) p_{xy|n}^Y(w_{\cdot y})  - \lambda_{ny}^Y p_{xy}^Y(w_{\cdot y}) < 1 \text{ for } x \in \mathcal{B}_n^X.
\end{align*}
If we choose the set of scalars, $(c_{xy}^X,c_{xy}^Y)$ in Eq. \eqref{eq_FixedPointRobustMatch}, by applying the following rules
\begin{align*}
    c_{xy}^X 
    &= \sum_{k=1}^{|K|}\mathbb{1}(y \in \mathcal{B}_k^Y) \lambda_{xk}^X  \; \forall \; (x,y) \in \mathcal{X} \times \mathcal{Y},\\  
    c_{xy}^Y
    &=\sum_{\ell=1}^{|L|}\mathbb{1}(x \in \mathcal{B}_{\ell}^X) \lambda_{\ell y}^Y  \; \forall \; (x,y) \in \mathcal{X} \times \mathcal{Y},
\end{align*}
then Assumption \ref{assumption_Elasticity} is satisfied. Hence, iterating on Eq. \eqref{eq_algorithm} is ensured to converge to the unique distribution of equilibrium wages.

\subsection{The generalized nested logit case}
Let there exist $|K|$ and $|L|$ nests of firms and workers. Each firm and worker can belong to each nest with a varying degree. Let this degree that be denoted by $\alpha_{xyk}^Y \geq 0$ and $\alpha_{xyk}^X \geq 0$, where it must hold that $\sum_{k=1}^K \alpha_{xyk}^X = 1$ and $\sum_{\ell=1}^L \alpha_{\ell xy}^Y = 1$. The generalized nested logit choice probabilities for workers and firms are given as
\begin{align*}
    p_{xy}^X(w_{x \cdot}) 
    &= \frac{\sum_{m=1}^{|K|}\alpha_{ym}^X\exp{\left(\Tilde{v}_{xy}^X / \lambda_{xm}^X \right)} \left[\sum_{j \in \mathcal{Y}} \alpha_{jm}^X \exp{\left( \Tilde{v}_{xj}^X / \lambda_{xm}^X \right)} \right]^{\lambda_{xm}^X - 1}}{1 + \sum_{k=1}^{|K|} \left[\sum_{j \in \mathcal{Y}} \alpha_{jk}^X \exp{\left( \Tilde{v}_{xj}^X / \lambda_{xk}^X \right)} \right]^{\lambda_{xk}^X}} \text{ for } y \in \mathcal{Y}, \\
    p_{xy}^Y(w_{\cdot y}) 
    &=  \frac{\sum_{n=1}^{|L|}\alpha_{nx}^Y\exp{\left(\Tilde{v}_{xy}^Y / \lambda_{ny}^Y \right)} \left[\sum_{i \in \mathcal{X}} \alpha_{ix}^Y \exp{\left( \Tilde{v}_{iy}^Y / \lambda_{ny}^Y \right)} \right]^{\lambda_{ny}^Y - 1}}{1 + \sum_{\ell=1}^{|L|} \left[\sum_{i \in \mathcal{X}}  \alpha_{i \ell}^Y \exp{\left( \Tilde{v}_{iy}^Y / \lambda_{\ell y}^Y \right)} \right]^{\lambda_{\ell y}^Y}} \text{ for } x \in \mathcal{X},
\end{align*} 
\cite{nielsen2021iterative} showed that the product of the smallest nesting parameter and the own-elasticity is guaranteed to be less than one. Therefore, we can ensure the convergence of fixed-point iterations by setting the scalars according to the following rules
\begin{align*}
    c_{xy}^X 
    &= \min_{k \in (1, \cdots, |K|)} \lambda_{xk}^X  \; \forall \; (x,y) \in \mathcal{X} \times \mathcal{Y},\\  
    c_{xy}^Y
    &=\min_{\ell \in (1, \cdots, |L|)} \lambda_{\ell y}^Y  \; \forall \; (x,y) \in \mathcal{X} \times \mathcal{Y}.
\end{align*}

\section{Conclusion}\label{sec_conclude}
In this note, we propose an easy-to-implement algorithm for solving one-to-one matching models with transferable utility. The algorithm is based on fixed-point iterations, where the guess for the equilibrium transfers is iteratively updated until demand and supply equate, and it is ensured to converge when the own-elasticities of the optimal choice probabilities are bounded from above. 

We hope that our results will facilitate further empirical research on matching markets by (1) treating demand and supply within a unified structural framework and (2) addressing the endogeneity of equilibrium transfers (such as wages) in a manner similar to the pioneering work of \cite{BLP1995}, who introduced a widely adopted empirical framework for estimating discrete-choice demand models that tackles price endogeneity through a combination of instrumental variables and a logit random coefficients model.

\newpage
\appendix
\section{Derivation of the system of fixed-point equations}\label{sec_derivation}
\cite{mcfadden1978nested} showed that when the stochastic payoffs follow any generalized extreme value (GEV) distribution, the resulting choice probabilities can be written in terms of the choice probability generating functions
, $g_x^X: \mathbb{R}^{|\mathcal{Y}|+1} \rightarrow \mathbb{R}$ and $g_y^Y: \mathbb{R}^{|\mathcal{X}|+1} \rightarrow \mathbb{R}$.
\begin{align}
    p_{xy}^X(w_{x \cdot}) = \frac{\nabla_{\exp{\left(\Tilde{v}_{xy}^X\right)}} g_x^X\left(\exp{\left(\Tilde{v}_{x \cdot}^X\right)}\right) \exp{\left(\Tilde{v}_{xy}^X\right)}}{g_x^X\left(\exp{\left(\Tilde{v}_{x \cdot}^X\right)}\right)} \; \forall \; y \in \mathcal{Y}_0, \label{eq_GEV_CCP_worker}\\
    p_{xy}^Y(w_{\cdot y}) = \frac{\nabla_{\exp{\left(\Tilde{v}_{xy}^Y\right)}} g_y^Y\left(\exp{\left(\Tilde{v}_{\cdot y}^Y\right)}\right) \exp{\left(\Tilde{v}_{xy}^Y\right)}}{g_y^Y\left(\exp{\left(\Tilde{v}_{\cdot y}^Y\right)}\right)} \; \forall \; x \in \mathcal{X}_0. \label{eq_GEV_CCP_firm}
\end{align}
To simplify notation, define the functions $G_x^X: \mathbb{R}^{|\mathcal{Y}|} \rightarrow \mathbb{R}$ and $G_y^Y: \mathbb{R}^{|\mathcal{X}|} \rightarrow \mathbb{R}$
\begin{align}
    G_x^X(w_{x \cdot}) 
    &= \log g_x^X\left(\exp{\left(\Tilde{v}_{x \cdot}^X\right)}\right) \; \forall \; x \in \mathcal{X}, \label{eq_Surplus_worker} \\ 
    G_y^Y(w_{\cdot y}) 
    &= \log g_y^Y\left(\exp{\left(\Tilde{v}_{\cdot y}^Y\right)}\right) \; \forall \; y \in \mathcal{Y}. \label{eq_Surplus_firm}
\end{align}
Next, take the logarithm of the market-clearing condition, given by Eq. \eqref{eq_market_clearing}, insert the choice probabilities, given by Eq. \eqref{eq_GEV_CCP_worker}-\eqref{eq_GEV_CCP_firm}, the deterministic payoffs, given by Eq. \eqref{eq_payoff_workers} and \eqref{eq_payoff_firms}, and rearrange
\begin{align}
    w_{xy}^*
    = \tfrac{\sigma_x^X \sigma_y^Y}{\sigma_x^X + \sigma_y^Y} \bigg[
    &\log n_y^Y + \tfrac{\beta_{xy}^Y}{\sigma_y^Y} + \log \nabla_{\exp{\left(\Tilde{v}_{xy}^Y\right)}} g_y^Y\left(\exp{\left(\Tilde{v}_{\cdot y}^Y\right)}\right) - G_y^Y(w_{\cdot y}^*) \nonumber \\
    -&\log n_x^X - \tfrac{\beta_{xy}^X}{\sigma_x^X} - \log \nabla_{\exp{\left(\Tilde{v}_{xy}^X\right)}} g_x^X\left(\exp{\left(\Tilde{v}_{x \cdot}^X\right)}\right) + G_x^X(w_{x \cdot}^*) \bigg]. \label{eq_FixedPointComplicated}
\end{align}
Eq. \eqref{eq_FixedPointComplicated} can be seen as a generalization from EV1 to GEV of the one proposed by \cite{dupuy2022note}.

The system of fixed-point equations described by Eq. \eqref{eq_FixedPointComplicated} can be simplified by first adding and subtracting the choice probabilities of the workers and firms, and then inserting the expressions for the choice probabilities and the deterministic payoffs
\begin{align*}
    w_{xy}^*
    = \tfrac{\sigma_x^X \sigma_y^Y}{\sigma_x^X + \sigma_y^Y} \bigg[
    &\log n_y^Y + \tfrac{\beta_{xy}^Y}{\sigma_y^Y} + \log \nabla_{\exp{\left(\Tilde{v}_{xy}^Y\right)}} g_y^Y\left(\exp{\left(\Tilde{v}_{\cdot y}^Y\right)}\right) - G_y^Y(w_{\cdot y}^*) \\
    +& \log p_{xy}^Y(w_{\cdot y}^*) - \left( \tfrac{\beta_{xy}^Y}{\sigma_y^Y} - \tfrac{w_{xy}^*}{\sigma_y^Y} + \log \nabla_{\exp{\left(\Tilde{v}_{xy}^Y\right)}} g_y^Y\left(\exp{\left(\Tilde{v}_{\cdot y}^Y\right)}\right) - G_y^Y(w_{\cdot y}^*) \right) \\
    -&\log n_x^X - \tfrac{\beta_{xy}^X}{\sigma_x^X} - \log \nabla_{\exp{\left(\Tilde{v}_{xy}^X\right)}} g_x^X\left(\exp{\left(\Tilde{v}_{x \cdot}^X\right)}\right) + G_x^X(w_{x \cdot}^*) \\
    +& \left( \tfrac{\beta_{xy}^X}{\sigma_x^X} + \tfrac{w_{xy}^*}{\sigma_x^X} - \log \nabla_{\exp{\left(\Tilde{v}_{xy}^X\right)}} g_x^X\left(\exp{\left(\Tilde{v}_{x \cdot}^X\right)}\right) + G_x^X(w_{x \cdot}^*) \right) - \log p_{xy}^X(w_{x \cdot}^*) \bigg].
\end{align*}
It follows that most of the terms cancel out, and we are left with Eq. \eqref{eq_FixedPoint}.

\section{Proof of Theorem \ref{thm_robust}}\label{sec_proof}
Let $\nabla_W F$ denote the Jacobian of $F$ with respect to the vector of wages, $W$. The infinity norm of $\nabla_W F$ can then be expressed as
\begin{align*}
    \lVert \nabla_W F(W) \lVert_{\infty} 
    &= \text{max}_{x,y} \: \left\{\sum_{j=1}^{|\mathcal{Y}|}\sum_{i=1}^{|\mathcal{X}|} \left|\nabla_{w_{ij}} f_{xy}(W)\right|\right\},
\end{align*}
By applying the mean value theorem, $F$ can be shown to define a contraction, if the infinity norm of the Jacobian is less than one
\begin{align*}
    \lVert F(W_1) - F(W_0) \rVert_{\infty} 
    &= \lVert\nabla F(W) \cdot (W_1 - W_0) \rVert_{\infty} \\
    &\leq \lVert  \nabla F(W) \rVert_{\infty} \cdot \lVert W_1 - W_0 \rVert_{\infty}.
\end{align*}

Under Assumption \ref{assumption_Elasticity}, for any combination of $(i,j,x,y)$ the functions $\nabla_{ij}k_{xy}^X: \mathbb{R}^{|\mathcal{Y}|} \rightarrow \mathbb{R}$ and $\nabla_{ij}k_{xy}^Y: \mathbb{R}^{|\mathcal{X}|} \rightarrow \mathbb{R}$ must exist
\begin{align}
    c_{xy}^X \frac{\nabla_{\Tilde{v}_{ij}^X} p_{xy}^X(w_{x \cdot})}{p_{xy}^X(w_{x \cdot})} 
    &= \mathbb{1}(i=x,j=y) - \mathbb{1}(i=x) \nabla_{ij}k_{xy}^X(w_{x \cdot}), \label{eq_robust_elasticity_worker}\\
    c_{xy}^Y \frac{\nabla_{\Tilde{v}_{ij}^Y} p_{xy}^Y(w_{\cdot y})}{p_{xy}^Y(w_{\cdot y})}
    &= \mathbb{1}(i=x,j=y) - \mathbb{1}(j=y) \nabla_{ij}k_{xy}^Y(w_{\cdot y}). \label{eq_robust_elasticity_firm}
\end{align}
In any additive random utility model alternatives are restricted to be substitutes. Under Assumption \ref{assumption_Elasticity}, this implies that $\nabla_{ij}k_{xy}^X$ and $\nabla_{ij}k_{xy}^Y$ must be positive for any $w_{x \cdot} \in \mathbb{R}^{|\mathcal{Y}|}$ and $w_{x \cdot} \in \mathbb{R}^{|\mathcal{X}|}$
\begin{align}
    \nabla_{xj}k_{xy}^X(w_{x \cdot}) > 0 \; \forall \; j \in \mathcal{Y}_0, \label{eq_property1}\\ 
    \nabla_{iy}k_{xy}^Y(w_{\cdot y}) > 0 \; \forall \; i \in \mathcal{X}_0, \label{eq_property2}
\end{align}
and sum to unity
\begin{align}
    1 - \nabla_{xy}k_{xy}^X(w_{x \cdot}) 
    &= \sum_{j \neq y}^{|\mathcal{Y}|} \nabla_{xj}k_{xy}^X(w_{x \cdot})  \Leftrightarrow 1 = \sum_{j=0}^{|\mathcal{Y}|} \nabla_{xj}k_{xy}^X(w_{x \cdot}), \label{eq_property3} \\ 
    1 - \nabla_{xy}k_{xy}^Y(w_{\cdot y}) 
    &= \sum_{i \neq x}^{|\mathcal{X}|} \nabla_{iy}k_{xy}^Y(w_{\cdot y})  \Leftrightarrow 1 = \sum_{i=0}^{|\mathcal{X}|} \nabla_{iy}k_{xy}^Y(w_{\cdot y}). \label{eq_property4}
\end{align}
The derivative of $f_{xy}$ with respect to $w_{ij}$ is for any combination of $(i,j,x,y)$ given in terms of the elasticities of the choice probabilities
\small
\begin{align*}
    \nabla_{w_{ij}} f_{xy}(W) 
    &= \mathbb{1}(i=x, j=y) - \tfrac{c_{xy}^Y\sigma_y^Y}{c_{xy}^X\sigma_x^X + c_{xy}^Y\sigma_x^Y} c_{xy}^X \frac{\nabla_{\Tilde{v}_{ij}^X} p_{xy}^X(w_{x \cdot})}{p_{xy}^X(w_{x \cdot})} - \tfrac{c_{xy}^X\sigma_x^X}{c_{xy}^X\sigma_x^X + c_{xy}^Y\sigma_x^Y} c_{xy}^Y \frac{\nabla_{\Tilde{v}_{ij}^Y} p_{xy}^Y(w_{\cdot y})}{p_{xy}^Y(w_{\cdot y})}.
\end{align*}
\normalsize
Note that $\nabla_{\Tilde{v}_{ij}^X} p_{xy}^X(w_{x \cdot})=0$ when $i \neq x$, and $\nabla_{\Tilde{v}_{ij}^Y} p_{xy}^Y(w_{\cdot y})=0$ when $j \neq y$. Next, insert Eq. \eqref{eq_robust_elasticity_worker} and \eqref{eq_robust_elasticity_firm} to obtain a simplified expression for $\nabla_{w_{ij}} f_{xy}$
\begin{align*}
    \nabla_{w_{ij}} f_{xy}(W) 
    &= \tfrac{c_{xy}^Y\sigma_y^Y}{c_{xy}^X\sigma_x^X + c_{xy}^Y\sigma_x^Y} \mathbb{1}(i=x)\nabla_{ij}k_{xy}^X(w_{x \cdot}) + \tfrac{c_{xy}^X\sigma_x^X}{c_{xy}^X\sigma_x^X + c_{xy}^Y\sigma_x^Y}\mathbb{1}(j=y)\nabla_{ij}k_{xy}^Y(w_{\cdot y}),
\end{align*}
Using the properties of $\nabla_{ij} k_{xy}^X$ and $\nabla_{ij} k_{xy}^Y$ described in Eq. \eqref{eq_property1}-\eqref{eq_property4}, it can be shown that the infinity norm of the Jacobian of $F$ is strictly less than one, i.e., $\lVert \nabla_W F(W) \rVert_{\infty} < 1$, due to the existence of outside options
\small
\begin{align*}
    \lVert \nabla_W F(W) \lVert_{\infty} 
    &= \text{max}_{x,y} \: \bigg\{\tfrac{c_{xy}^Y\sigma_y^Y}{c_{xy}^X\sigma_x^X + c_{xy}^Y\sigma_x^Y} \sum_{j=1}^{|\mathcal{Y}|} \nabla_{xj}k_{xy}^X(w_{x \cdot}) + \tfrac{c_{xy}^X\sigma_x^X}{c_{xy}^X\sigma_x^X + c_{xy}^Y\sigma_x^Y} \sum_{i=1}^{|\mathcal{X}|} \nabla_{iy}k_{xy}^Y(w_{\cdot y}) \bigg\}\\
    &= \text{max}_{x,y} \: \bigg\{\tfrac{c_{xy}^Y\sigma_y^Y}{c_{xy}^X\sigma_x^X + c_{xy}^Y\sigma_x^Y} \bigg[1-\nabla_{x0}k_{xy}^X(w_{x \cdot})\bigg] + \tfrac{c_{xy}^X\sigma_x^X}{c_{xy}^X\sigma_x^X + c_{xy}^Y\sigma_x^Y}\bigg[1 - \nabla_{0y}k_{xy}^Y(w_{\cdot y})\bigg] \bigg\}\\ 
    &= \text{max}_{x,y} \: \bigg\{1 - \tfrac{c_{xy}^Y\sigma_y^Y}{c_{xy}^X\sigma_x^X + c_{xy}^Y\sigma_x^Y} \nabla_{x0}k_{xy}^X(w_{x \cdot}) - \tfrac{c_{xy}^X\sigma_x^X}{c_{xy}^X\sigma_x^X + c_{xy}^Y\sigma_x^Y} \nabla_{0y}k_{xy}^Y(w_{\cdot y}) \bigg\}.
\end{align*}
\normalsize



\newpage
\bibliography{bibtex.bib}

\begin{thebibliography}{22}
\providecommand{\natexlab}[1]{#1}

\bibitem[{Andersen and Lee(2022)}]{andersenlee2022}
\textsc{Andersen, E.~S.} and \textsc{Lee, Y.~J.} (2022). Multidimensional
  matching and labor market complementarity. In \textit{Essays on Structural
  Microeconometrics. Perturbed utility and equilibrium models}, phd thesis
  \textit{chp. 1}, Department of Economics, University of Copenhagen, pp.
  1--48.

\bibitem[{Bajari and Fox(2013)}]{bajari2013estimating}
\textsc{Bajari, P.} and \textsc{Fox, J.~T.} (2013). Estimating dynamic models
  of imperfect competition. \textit{Econometrica}, \textbf{81}~(3), 989--1019.

\bibitem[{Berry \textit{et~al.}(1995)Berry, Levinsohn and Pakes}]{BLP1995}
\textsc{Berry, S.}, \textsc{Levinsohn, J.} and \textsc{Pakes, A.} (1995).
  Automobile prices in market equilibrium. \textit{Econometrica},
  \textbf{63}~(4), 841--890.

\bibitem[{Chiappori \textit{et~al.}(2010)Chiappori, McCann and
  Nesheim}]{chiappori2010hedonic}
\textsc{Chiappori, P.-A.}, \textsc{McCann, R.~J.} and \textsc{Nesheim, L.}
  (2010). Hedonic pricing in matching models. \textit{Journal of Political
  Economy}, \textbf{118}~(6), 1476--1508.

\bibitem[{Chiappori \textit{et~al.}(2017)Chiappori, Salanié and
  Weiss}]{ChiapporiEtAl2017}
\textsc{---}, \textsc{Salanié, B.} and \textsc{Weiss, Y.} (2017). Partner
  choice, investment in children, and the marital college premium.
  \textit{American Economic Review}, \textbf{107}~(8), 2109–67.

\bibitem[{Choo and Siow(2006)}]{choo2006marriage}
\textsc{Choo, E.} and \textsc{Siow, A.} (2006). Who marries whom and why.
  \textit{Journal of Political Economy}, \textbf{114}~(1), 175--201.

\bibitem[{Costinot and Vogel(2015)}]{costinot2015}
\textsc{Costinot, A.} and \textsc{Vogel, J.} (2015). Beyond ricardian trade and
  comparative advantage. \textit{Annual Review of Economics}, \textbf{7},
  31--62.

\bibitem[{Dubé \textit{et~al.}(2012)Dubé, Fox and Su}]{Dube2012}
\textsc{Dubé, J.}, \textsc{Fox, J.~T.} and \textsc{Su, C.} (2012). {Improving
  the Numerical Performance of Static and Dynamic Aggregate Discrete Choice
  Random Coefficients Demand Estimation}. \textit{Econometrica},
  \textbf{80}~(5), 2231--2267.

\bibitem[{Dupuy and Galichon(2014)}]{dupuy2014JPE}
\textsc{Dupuy, A.} and \textsc{Galichon, A.} (2014). Personality traits and the
  marriage market. \textit{Journal of Political Economy}, \textbf{122}~(6),
  1271--1319.

\bibitem[{Dupuy and Galichon(2022)}]{dupuy2022note}
\textsc{---} and \textsc{---} (2022). A note on the estimation of job amenities
  and labor productivity. \textit{Quantitative Economics}, \textbf{13}~(3),
  1063--1081.

\bibitem[{Dupuy \textit{et~al.}(2020)Dupuy, Galichon, Jaffe and
  Kominers}]{DupuyTaxation2020}
\textsc{---}, \textsc{---}, \textsc{Jaffe, S.} and \textsc{Kominers, S.~D.}
  (2020). Taxation in matching markets. \textit{International Economic Review},
  \textbf{61}~(4), 1591--1634.

\bibitem[{Fox(2018)}]{fox2018}
\textsc{Fox, J.~T.} (2018). Estimating matching games with transfers.
  \textit{Quantitative Economics}, \textbf{9}~(1), 1--38.

\bibitem[{Fox \textit{et~al.}(2018)Fox, Yang and Hsu}]{fox2018unobserved}
\textsc{---}, \textsc{Yang, C.} and \textsc{Hsu, Y.-W.~L.} (2018). Unobserved
  heterogeneity in matching games. \textit{Journal of Political Economy},
  \textbf{126}~(5), 1793--1842.

\bibitem[{Gabaix and Landier(2008)}]{gabaix2008}
\textsc{Gabaix, X.} and \textsc{Landier, A.} (2008). Why has ceo pay increased
  so much? \textit{The Quarterly Journal of Economics}, \textbf{123}~(1),
  49--100.

\bibitem[{Galichon and Salani{\'e}(2017)}]{galichon2017iia}
\textsc{Galichon, A.} and \textsc{Salani{\'e}, B.} (2017). Iia in separable
  matching markets, unpublished.

\bibitem[{Galichon and Salanié(2022)}]{galichon2023cupidsinvisiblehandsocial}
\textsc{---} and \textsc{Salanié, B.} (2022). Cupid's invisible hand: Social
  surplus and identification in matching models. \textit{The Review of Economic
  Studies}, \textbf{89}~(5), 2600–2629.

\bibitem[{Grigolon and Verboven(2014)}]{grigolon2014nestedRC}
\textsc{Grigolon, L.} and \textsc{Verboven, F.} (2014). Nested logit or random
  coefficients logit? a comparison of alternative discrete choice models of
  product differentiation. \textit{The Review of Economics and Statistics},
  \textbf{96}~(5), 916--935.

\bibitem[{Knittel and Metaxoglou(2014)}]{Knittel2014}
\textsc{Knittel, C.~R.} and \textsc{Metaxoglou, K.} (2014). {Estimation of
  Random-Coefficient Demand Models: Two Empiricists' Perspective}. \textit{The
  Review of Economics and Statistics}, \textbf{96}~(1), 34--59.

\bibitem[{Lee and Seo(2016)}]{Lee2016}
\textsc{Lee, J.} and \textsc{Seo, K.} (2016). {Revisiting the nested
  fixed-point algorithm in BLP random coefficients demand estimation}.
  \textit{Economics Letters}, \textbf{149}~(C), 67--70.

\bibitem[{McFadden(1978)}]{mcfadden1978nested}
\textsc{McFadden, D.} (1978). Modelling the choice of residential location. In
  A.~Karlqvist, L.~Lundqvist, F.~Snickars and J.~W. Weibull (eds.),
  \textit{Spatial Interaction Theory and Planning Models}, Amsterdam:
  North-Holland, pp. 75--96.

\bibitem[{Nielsen(2021)}]{nielsen2021iterative}
\textsc{Nielsen, N.} (2021). Iterative ccp estimation beyond the linear logit
  model. In \textit{The Econometrics of Perturbed Utility and Demand
  Inversion}, phd thesis \textit{chp. 3}, Department of Economics, University
  of Copenhagen, pp. 89--118.

\bibitem[{Tervio(2008)}]{tervio2008}
\textsc{Tervio, M.} (2008). The difference that ceos make: An assignment model
  approach. \textit{American Economic Review}, \textbf{98}~(3), 642--668.

\end{thebibliography}
\addcontentsline{toc}{chapter}{References}
\bibliographystyle{ecca}

\end{document}